\documentclass[conference]{IEEEtran}
\IEEEoverridecommandlockouts
\usepackage{cite}
\usepackage[table]{xcolor}
\usepackage{amsmath,amssymb,amsfonts}
\usepackage{algorithmic}
\usepackage{graphicx}
\usepackage{makecell}
\usepackage{textcomp}
\usepackage{xcolor}
\usepackage[most]{tcolorbox}
\usepackage{balance}
\def\BibTeX{{\rm B\kern-.05em{\sc i\kern-.025em b}\kern-.08em
    T\kern-.1667em\lower.7ex\hbox{E}\kern-.125emX}}
\tcbset{textmarker/.style={
        enhanced,
        parbox=false,boxrule=0mm,boxsep=0mm,arc=0mm,
        outer arc=0mm,left=5mm,right=3mm,top=7pt,bottom=7pt,
        toptitle=1mm,bottomtitle=1mm}}
\newtcolorbox{noteBox}{textmarker,
    borderline west={4pt}{0pt}{gray},
    colback=gray!10!white}
\newcommand{\note}[1]{\begin{noteBox} #1 \end{noteBox}}
    
\begin{document}

\title{Assessing Teamwork Dynamics in Software Development Projects}

\author{
    \IEEEauthorblockN{Santiago Berrezueta-Guzman}
    \IEEEauthorblockA{
        \textit{Technical University of Munich}\\
        Heilbronn, Germany \\
        s.berrezueta@tum.de}
    \and
    \IEEEauthorblockN{Ivan Parmacli}
    \IEEEauthorblockA{
        \textit{Technical University of Munich}\\
        Heilbronn, Germany \\
        ivan.parmacli@tum.de}
    \and
    \IEEEauthorblockN{Mohammad Kasra Habib}
    \IEEEauthorblockA{
        \textit{Technical University of Munich}\\
        Heilbronn, Germany \\
        kasra.habib@tum.de}
    \and
    \hspace*{4.5cm}\IEEEauthorblockN{
        Stephan Krusche \hspace{3cm} Stefan Wagner
    }
    \IEEEauthorblockA{
        \hspace*{4.6cm}\textit{Technical University of Munich}
        \hspace*{1cm}\textit{Technical University of Munich}\\
        \hspace*{4.9cm}Munich, Germany \hspace{2.9cm} Heilbronn, Germany \\
        \hspace*{5.2cm}krusche@tum.de \hspace{2.8cm} stefan.wagner@tum.de
    }
}

\maketitle

\begin{abstract}
This study investigates teamwork dynamics in student software development projects through a mixed-method approach combining quantitative analysis of GitLab commit logs and qualitative survey data. We analyzed individual contributions across six project phases, comparing self-reported and actual contributions to measure discrepancies. Additionally, a survey captured insights on team leadership, conflict resolution, communication practices, and workload perceptions. Findings reveal that teams with minimal contribution discrepancies achieved higher project grades and exam pass rates. In contrast, teams with more significant discrepancies experienced lower performance, potentially due to role clarity and communication issues. These results underscore the value of shared leadership, structured conflict resolution, and regular feedback in fostering effective teamwork, offering educators strategies to enhance collaboration in software engineering education through self-reflection and balanced workload allocation.

\end{abstract}


\section{Introduction}
Teamwork remains a cornerstone of effective software education, equipping students with essential collaboration skills that mirror real-world development environments. As software development increasingly relies on agile methodologies and cross-functional teams, working collaboratively is crucial for success in the field \cite{stray2016daily}. Collaborative learning in software education enhances technical skills and fosters the soft skills necessary for professional communication, adaptability, and problem-solving within a team context. Research indicates that students working in teams are better prepared to handle complex software challenges, as they benefit from diverse perspectives and collective problem-solving efforts, which enrich their understanding of software practices \cite{sancho2009learning}. 

These collaborative experiences also increase motivation and engagement as students learn to navigate group dynamics and resolve conflicts, ultimately strengthening their preparedness for industry demands \cite{whitehead2007collaboration, whitehead2010collaborative}. Studies suggest teamwork fosters a deeper understanding of software development processes because students encounter and overcome challenges that individual work might not present \cite{hanks2011pair}.

This study examines the dynamics of teamwork in a semester-long software development project, divided into six incremental deliverables. It explores how these dynamics influence team performance, collaboration quality, and academic outcomes. The importance of teamwork in software engineering education is emphasized mainly through the lens of two key research questions (RQ). 

\textbf{RQ1}: How do teamwork dynamics influence team performance and collaboration? 

\textbf{RQ2}: What teamwork factors most significantly impact academic success?

This paper introduces the topic and its significance in software engineering education. Section \ref{RW} presents related work, situating this study within existing research. Section \ref{M} describes the methodology, while Section \ref{R} outlines the results. Section \ref{D} offers an in-depth discussion of the findings, and Section \ref{C} concludes with insights and implications drawn from the study.

\section{Related Work}\label{RW}

In recent years, collaborative learning has gained prominence in educational frameworks, particularly in software engineering, where teamwork is integral to professional success \cite{francescato2006evaluation, sumtsova2018collaborative}. Effective collaboration fosters critical interpersonal skills, enhances problem-solving capabilities, and mirrors the team-based dynamics of industry environments. As universities increasingly adopt cooperative methodologies, studies reveal the profound impact of structured teamwork on skill development, preparation for real-world settings, and overall student engagement \cite{roselli2016collaborative}. This section reviews prior research on cooperative learning models and their effectiveness in cultivating the social and technical skills essential for future professionals.

Chen et al. introduce a novel approach to assessing teamwork skills in software engineering education, emphasizing collaborative learning and active participation. This approach, applied in various software engineering courses, assesses team performance and individual contributions, encouraging students to engage deeply in team activities. By structuring teams based on learning styles and providing clear roles and assessment metrics, the model helps balance knowledge distribution within teams. The study underscores the importance of teamwork in software engineering and addresses common challenges, such as fairness in peer evaluation, to promote better team dynamics and skills development in educational or even professional settings \cite{chen2011assessing}.

Mendo-Lázaro et al. examined the effectiveness of cooperative learning for developing teamwork-related social skills among university students. Using a quasi-experimental design, they compare control and experimental groups to analyze skills like self-assertion, information reception, and information dissemination changes. The study found that cooperative learning significantly enhances these skills, especially with prolonged engagement and structured group setups. This work highlights the role of collaborative learning in preparing students for professional collaboration, suggesting it as an essential approach for building interpersonal skills, which are crucial for effective teamwork in higher education contexts \cite{mendo2018cooperative}. 

Raibulet and Fontana explore the impact of collaborative and teamwork-oriented software development within an undergraduate software engineering course. Their study, conducted with third-year computer science students, integrates real-world tools such as GitHub, SonarQube, and Microsoft Project to simulate industry practices. Through GitHub, students manage source code collaboratively, SonarQube enables quality assessment, and Microsoft Project supports project management. Feedback reveals that students found these tools beneficial for enhancing team coordination, task management, and code quality awareness while highlighting challenges in using industry-level tools in academic settings. This study underscores the value of introducing professional tools and collaborative methods in academic courses to bridge the gap between educational practices and industry needs \cite{raibulet2018collaborative}.

In previous work \cite{10837659, 10734012}, we analyzed teamwork dynamics among first-semester programming students in a structured project setting to uncover patterns affecting team success and individual engagement. The study uses GitLab logs and surveys to capture contributions and estimate errors, revealing that students tend to overestimate their involvement, a bias influenced by visibility and interpretation of roles. Notably, nationality and gender did not impact team outcomes, while “lone wolf” students frequently struggled with collaboration, leading to team dropouts. Findings emphasize the importance of active participation, as students who contributed more significantly performed better in presentations and exams and avoided plagiarism misconducts, suggesting that collaborative projects can enhance technical and interpersonal skills essential for professional settings \cite{10229422}. 

\section{Methodology}\label{M}

To comprehensively evaluate teamwork dynamics and individual contributions in a collaborative software development project, we adopted a mixed-method approach that combined quantitative and qualitative data.

\subsection{Objective Contribution and Discrepancy Analysis}

We analyzed the backlog of GitLab, tracking the number of commits for each team member in each incremental assignment and getting objective data on individual contributions. This allows for a direct comparison between self-reported contributions and actual workload data, highlighting discrepancies where applicable. 

Equation \ref{eq} calculates the percentage difference between a student's estimated contribution \( EC_i \) and their actual contribution \( RC_i \) relative to the total group effort. 

\begin{equation}
\text{Difference} = \left| EC_i - \frac{RC_i}{\left(\sum_{j=1}^{5} RC_j\right)} \right| \times 100 \% \label{eq}
\end{equation}

Specifically, \( EC_i \) represents the student’s self-assessed contribution, while \( RC_i \) is the measured contribution. The term \( \sum_{j=1}^{5} RC_j \) is the total contribution of all group members (assuming a group of five), making \( \frac{RC_i}{\sum_{j=1}^{5} RC_j} \) the student’s actual contribution as a proportion of the group’s total. By taking the absolute difference \( \left| EC_i - \frac{RC_i}{\sum_{j=1}^{5} RC_j} \right| \), we obtain the magnitude of discrepancy between estimated and actual contributions. Multiplying by 100 converts this difference into a percentage, measuring how closely a student’s self-assessment aligns with their measured contribution, where larger values indicate more significant discrepancies.

\subsection{Assessment of Team Dynamics and Collaboration}

Secondly, we conducted a final survey to assess team dynamics, leadership roles, conflict resolution, communication strategies, and perceptions of workload distribution. Teams were asked to identify their team leader, explain the choice, evaluate whether they would work with the same team in future projects, and share how conflicts were managed. The survey also gathered insights on communication methods, the perceived fairness of workload distribution, and suggested improvements to team collaboration. Finally, participants rated their overall team experience, providing a comprehensive view of the team’s collaborative processes and areas for potential improvement.

\subsection{Project Description}

This project comprises six incremental assignments (IA) designed to simulate a real-world software development process. The objective is to develop a library management system, providing hands-on experience in core software engineering practices, including requirements elicitation, test case design, software development, maintenance, and quality assurance. At the end of the semester, the students present the completed project, demonstrating their understanding and application of these practices in a collaborative development environment.

\textbf{IA1. Requirements.} Students analyze and clarify system requirements by drafting specific questions for a library client, which will refine and guide the development of a command-line library management system that handles core processes, constraints, and reporting needs efficiently.

\textbf{IA2. Specification.} Instructors act as the customers, providing input as needed, such as maintenance instructions such as "IA5." They offered students the chance to adjust and implement new functionalities based on these inputs. Students developed a comprehensive Software Requirements Specification (SRS) document using this feedback and the initial requirements. This document included a domain model, use case diagrams, and tables outlining the core functionalities of the library management system. Additionally, they develop a prototype of the system's command-line interface (CLI) in Java to demonstrate user interaction. The SRS and prototype are submitted to GitLab, with clear documentation and instructions, to provide the client with an understanding of the system's structure and basic operation.

\textbf{IA3. Testing.} Students initiate the core implementation of the library management system, enhancing the UI prototype with domain model classes and in-memory data containers. They create unit tests for essential processes using JUnit 5, ensuring coverage of constraints and edge cases. A test coverage report, generated with tools like JaCoCo, highlights coverage metrics and insights. Compilation, testing, and coverage instructions are added to the README, and all components are submitted to GitLab.

\textbf{IA4. Implementation} Students implement the library management system, applying CI and code reviews to ensure quality. They create a branching model, use Maven or Gradle for builds, and develop core features like data import, searching, borrowing, and reporting, accessible via CLI. Through Test-Driven Development, they aim for 70\% test coverage, documenting processes and focusing on code quality and usability.

\textbf{IA5. Maintenance and Evolution.} Students evolve their library management system by adapting to new client requirements. They update the CSV import to track current borrowers, implement error handling, and update unit tests. Additionally, they create a new report that outputs the count and percentage of book copies per publisher, sorted alphabetically, to meet reporting needs. Quality is ensured through Merge Requests, code reviews, and extensive testing for all changes.

\textbf{IA6. Quality Assurance.} Students enhance the software’s maintainability by identifying and refactoring 16+ code vulnerabilities, focusing on aspects like naming, error handling, and SOLID principles. Each fix is documented in GitLab with separate Merge Requests, and a summary report details the vulnerabilities and improvements for final submission.

\section{Results}\label{R}

\subsection{Contribution and Discrepancy Analysis}

When we apply the equation \ref{eq} to each student in each group and then calculate the average difference among the members, we obtain the data presented in Table \ref{results}. We observe several instances where the \textit{Difference} between estimated (\textit{EC}) and real contribution (\textit{RC}) exceeds 20\%, as highlighted in red. A difference greater than 20\% suggests a significant discrepancy between a student’s self-assessed contribution (\( EC_i \)) and the measured contribution (\( RC_i \)) relative to the group's total contribution. This discrepancy indicates that, in these cases, students may have either overestimated or underestimated their contributions substantially.

The high differences (\textgreater 20\%) may point to several potential issues within those groups:
\begin{itemize}
    \item \textit{Misalignment in self-perception}: Students in these groups could have a distorted perception of their roles, thinking they contributed more or less than they actually did.
    \item \textit{Group dynamics}: High discrepancies can suggest issues with communication, uneven task distribution, or misunderstandings about each member’s roles and duties.
    \item \textit{Potential need for feedback}: These results could signal a need for more frequent feedback and assessment to ensure students develop a realistic understanding of their contributions and areas for improvement.
\end{itemize}

For example, groups like G13, G15, G16, and G20 have multiple high discrepancies, which may indicate persistent issues in self-assessment accuracy or possible coordination challenges in these teams.
Groups with isolated high discrepancies (such as G8 or G18) may have individual cases where specific students misjudged their contributions.

In general, a high percentage of team discrepancies can indicate group dynamics that need attention through improved communication, clearer role definitions, or enhanced self-assessment tools to help students better understand and reflect on their contributions.

\begin{table}[h!]
\centering
\caption{Difference between estimated and actual contribution (\%)}\label{results}
\begin{tabular}{|c|c|c|c|c|c|}
\hline
\multicolumn{6}{|c|}{\textbf{Difference Between Estimated and Actual Contribution}} \\
\hline
\textbf{Group} & \textbf{IA2(\%)} & \textbf{IA3(\%)} & \textbf{IA4(\%)} & \textbf{IA5(\%)} & \textbf{IA6(\%)} \\
\hline
G1 & 4.69 & 5.33 & 5.28 & 16.23 & 12.44 \\
G2 & 16.36 & 7.53 & 11.41 & 8.00 & 5.62 \\
G3 & 10.29 & 17.68 & 12.57 & 13.00 & 5.08 \\
G4 & 9.45 & 11.59 & 7.77 & 5.82 & 6.50 \\
G5 & \cellcolor{red!20}24.00 & 12.25 & 10.09 & \cellcolor{red!20}24.00 & 11.22 \\
G6 & \cellcolor{red!20}27.56 & \cellcolor{red!20}26.29 & 9.19 & 16.00 & 16.00 \\
G7 & \cellcolor{red!20}32.00 & \cellcolor{red!20}24.94 & 17.75 & 12.00 & 12.89 \\
G8 & \cellcolor{red!20}27.00 & 11.56 & 6.00 & \cellcolor{red!20}20.00 & 9.96 \\
G9 & 14.18 & 7.70 & 10.67 & 17.33 & 5.20 \\
G10 & 19.56 & 13.23 & 10.00 & \cellcolor{red!20}32.00 & \cellcolor{red!20}20.00 \\
G11 & 18.29 & 10.13 & 12.44 & 10.86 & 13.03 \\
G12 & 6.60 & 0.20 & 14.00 & 6.67 & \cellcolor{red!20}20.00 \\
G13 & \cellcolor{red!20}32.00 & 16.00 & 19.00 & \cellcolor{red!20}24.00 & \cellcolor{red!20}20.00 \\
G14 & \cellcolor{red!20}24.00 & \cellcolor{red!20}28.67 & 18.29 & 14.00 & 7.57 \\
G15 & \cellcolor{red!20}24.00 & \cellcolor{red!20}25.12 & \cellcolor{red!20}21.37 & \cellcolor{red!20}23.33 & \cellcolor{red!20}22.71 \\
G16 & \cellcolor{red!20}32.00 & \cellcolor{red!20}24.00 & 14.88 & \cellcolor{red!20}25.68 & 14.55 \\
G17 & 8.00 & 8.00 & 6.10 & 10.00 & 9.56 \\
G18 & \cellcolor{red!20}24.00 & 0.00 & \cellcolor{red!20}20.00 & \cellcolor{red!20}32.00 & 0.00 \\
G19 & \cellcolor{red!20}20.00 & 10.24 & \cellcolor{red!20}20.00 & 0.00 & 0.00 \\
G20 & \cellcolor{red!20}20.00 & \cellcolor{red!20}20.00 & \cellcolor{red!20}25.00 & 12.00 & \cellcolor{red!20}40.00 \\
G21 & \cellcolor{red!20}24.00 & 16.00 & 5.33 & \cellcolor{red!20}20.00 & 19.68 \\
G22 & 15.94 & 8.00 & 4.68 & 10.00 & 7.72 \\
G23 & 12.57 & 12.24 & 10.67 & 8.00 & \cellcolor{red!20}20.00 \\
\hline\hline

\end{tabular}
\end{table}

\begin{figure}[htbp]
\centerline{\includegraphics[width=7.4cm]{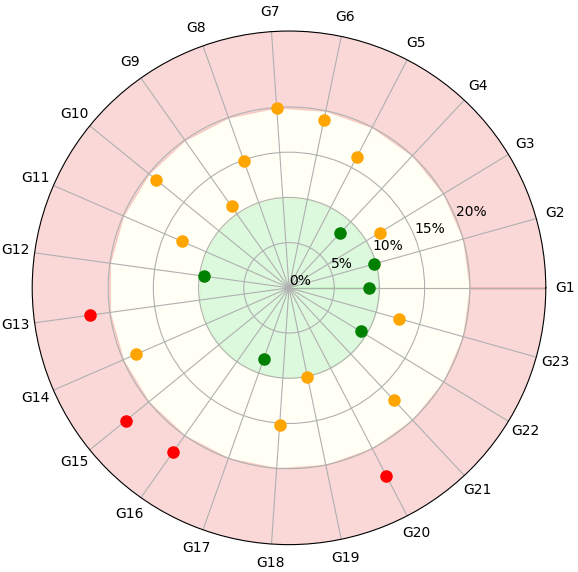}}
\caption{Distribution of the average difference between estimated and real contributions for each group in every assignment.}
\label{radar}
\end{figure}

Figure \ref{radar} displays the difference between estimated and accurate contributions for each group, with points plotted based on the magnitude of the discrepancy. Green points (differences \textless 10\%) indicate close alignment between estimated and actual contributions, orange points (10--20\%) show moderate discrepancies and red points (\textgreater 20\%) highlight significant mismatches. The background is color-coded to match these ranges: green for low, yellow for moderate, and red for high discrepancies. Groups like G5, G6, G7, G13, G14, G15, G16, and G20, which appear in the red zone, show significant gaps between self-assessed and actual contributions, as reflected in the table data. Groups with moderate differences, like G2, G3, and G9, fall in the yellow zone, while groups like G1, G4, G11, and G17, with close alignment, are in the green zone. This chart highlights groups with high discrepancies, suggesting where intervention might improve self-assessment accuracy and team dynamics.

Table \ref{grades} and the correlation analysis revealed a strong negative relationship (-0.83) between \textit{Average Difference (\%)} and \textit{Project Grade (\%)}, indicating that groups with higher average differences tend to have lower project grades. Additionally, a moderate positive correlation (0.53) exists between \textit{Project Grade (\%)} and \textit{Students Passed Exam}, suggesting that better project grades are associated with a higher number of students passing. Conversely, a moderate negative correlation (-0.62) between \textit{Average Difference (\%)} and \textit{Students Passed Exam} shows that groups with higher average differences tend to have fewer students passing. ANOVA results confirm statistically significant relationships among these variables, with p-values of \(1.73 \times 10^{-27}\) for \textit{Average Difference vs. Project Grade} and \(2.31 \times 10^{-32}\) for \textit{Project Grade vs. Students Passed Exam}.

\begin{table}[h!]
\centering
\caption{Performance analysis of student groups: comparison of average difference, project grade, and exam success rate}
\begin{tabular}{|c|c|c|c|}
\hline
\makecell{\textbf{Group}} & \makecell{\textbf{Average } \\ \textbf{Difference (\%)}} & \makecell{\textbf{Project } \\ \textbf{Grade (\%)}}& \makecell{\textbf{Students Passed} \\ \textbf{the Exam}} \\
\hline
G1  & \cellcolor{green!15} 8.80 & \cellcolor{green!15} 92.32 &  \cellcolor{green!15} 4/5 \\
G2  & \cellcolor{green!15} 9.78 & \cellcolor{green!15} 96.08 & \cellcolor{green!15} 3/5 \\ 
G3  & 11.72 & 82.98 & 3/5 \\
G4  & \cellcolor{green!15}8.23 & \cellcolor{green!15} 87.21 & \cellcolor{green!15} 4/5 \\
G5  & 16.31 & 79.54 & 4/5 \\
G6  & 19.01 & 61.35 & 2/5 \\
G7  & 19.92 & 81.39 & 3/5 \\
G8  & 14.90 & 78.45 & 2/5 \\
G9  & 11.02 & 81.86 & 3/5 \\
G10 & 18.96 & 79.98 & 2/5 \\
G11 & 12.95 & 76.82 & 2/5 \\
G12 & \cellcolor{green!15} 9.49  & \cellcolor{green!15} 90.09& \cellcolor{green!15} 4/5 \\
G13 & \cellcolor{red!20} 22.20& \cellcolor{red!15} 57.65 & \cellcolor{red!20} 1/5 \\
G14 & 18.50 & 71.23 & 1/5 \\
G15 & \cellcolor{red!20} 23.31 & \cellcolor{red!15} 76.54& \cellcolor{red!20} 1/5 \\
G16 & \cellcolor{red!20} 22.22 & \cellcolor{red!15} 57.54& \cellcolor{red!20} 1/5 \\
G17 & \cellcolor{green!15} 8.33 & \cellcolor{green!15} 94.33 & \cellcolor{green!15} 3/5 \\
G18 & 15.20 & 86.55 & 1/5 \\
G19 & 10.05 & 91.26 & 1/5 \\
G20 & \cellcolor{red!20} 23.40 & \cellcolor{red!15} 62.08& \cellcolor{red!20} 1/5 \\
G21 & 17.00 & 74.34 & 1/5 \\
G22 & \cellcolor{green!15} 9.27  & \cellcolor{green!15} 85.34& \cellcolor{green!15} 3/5 \\
G23 & 12.69 & 88.45 & 1/5 \\
\hline
\end{tabular}\label{grades}
\end{table}

\subsection{Assessment of Team Dynamics and Collaboration}

The survey results indicate that in this collaborative project, team leaders were typically chosen based on organizational skills, prior experience, or willingness to take on responsibilities. In contrast, some teams rotated leadership depending on members' expertise. Communication was facilitated through a mix of WhatsApp, Discord, Zoom, and in-person meetings, with teams generally meeting once or twice a week and scheduling additional sessions as deadlines approached. 

Conflict resolution involved open discussions, voting, and respect for each member's opinion, ensuring decisions aligned with project goals. However, most teams reported balanced workload distribution; three noted imbalances. Teams also suggested improvements, such as more precise timelines, structured agendas, enhanced commit practices, and adopting task-sharing platforms for better coordination. Overall, teams rated their experience from Good to Excellent, with the majority highlighting a positive, collaborative environment, though only one team expressed reluctance to work together again in the future. 

\section{Discussion}\label{D}

The results highlight essential correlations between self-assessment accuracy, team dynamics, and academic outcomes.

\note{\textbf{Finding 1 -- Alignment and Academic Success:} Teams with close alignment between self-reported and actual contributions performed better academically, with higher project grades and exam pass rates.}

This suggests that teams with accurate self-assessments and balanced contributions benefit from more effective collaboration, resulting in improved performance both in project work and individual assessments.

\note{\textbf{Finding 2 -- High Discrepancies and Team Issues:} Groups with high discrepancies in self-assessment often faced challenges like unclear roles, uneven task distribution, and weak communication, resulting in lower academic outcomes.}

Such issues may prevent some members from fully engaging with the work, leading to lower team performance and individual learning outcomes, e.g., G13, G15, G16, G20 show lower project grades and fewer students passing the exam.
Figure \ref{radar} and the highlighted cells in Table \ref{grades} reinforce these findings. These discrepancies indicate where interventions, such as fostering better communication practices or providing more precise role definitions, might enhance team cohesion and individual accountability. These insights underscore the value of tools and feedback mechanisms that promote self-awareness, equitable workload distribution, and realistic self-assessment among team members for educational settings.

\note{\textbf{Finding 3 -- Effective Leadership:} Teams with leaders chosen for organizational skills or expertise showed greater cohesion and resilience.}

Effective teamwork in educational settings relies on adequate leadership, open communication, and structured conflict resolution, with teams often assigning leadership based on organizational expertise. These findings align with Gren and Ralph’s \cite{gren2022makes} work on agile leadership, which emphasizes shared responsibility, team cohesion, and adaptability. Both studies suggest adequate leadership and a supportive team culture enhance educational or agile software development teamwork. In the same context, Modi and Strode's systematic review  \cite{modi2020leadership} of agile leadership aligns closely with these findings, mainly by emphasizing shared, situational, and transformational leadership styles that enable team collaboration. 

\note{\textbf{Finding 4 -- Communication and Conflict Resolution:} Open communication and structured conflict resolution supported effective teamwork and goal alignment.}

The findings reveal that teams primarily resolved conflicts through open discussions, aligning with the findings of Bobely et al. \cite{gobeli1998managing}, which emphasize constructive communication as a central tool for effective conflict resolution. Accurate self-assessment, equitable workload distribution, effective leadership, communication, and conflict resolution are linked to more robust team dynamics and performance.

\section{Conclusion}\label{C}

This study examines teamwork dynamics in student software development projects, identifying key factors influencing team success and academic outcomes. By comparing self-reported and actual contributions using GitLab commit data, we observed substantial discrepancies in certain groups, suggesting that some students overestimate or underestimate their contributions. Groups with minimal discrepancies and balanced contributions demonstrated stronger collaboration, higher project grades, and more students passing final exams. 

Findings emphasize the importance of adequate leadership, clear communication, and structured conflict resolution in educational teamwork settings. Teams that selected leaders based on organizational skills and communicated regularly showed greater cohesion and resilience. In contrast, teams with higher contribution discrepancies could benefit from more precise role definitions and equitable workload distribution.

Educators should implement regular feedback mechanisms and reflection exercises to encourage critical assessment of contributions and foster a realistic understanding of individual roles within a team. Promoting open discussions on workload allocation can prevent misunderstandings and support a productive team environment. Future research could explore the influence of leadership styles and conflict management approaches on team performance, contributing to best practices in software engineering education.

\balance
\bibliographystyle{ieeetr}
\bibliography{EDUCON_Paper}
\end{document}